\documentclass[a4paper,groupcitations]{article}
\usepackage{mwe}
\usepackage[T1]{fontenc}
\usepackage[ansinew]{inputenc}
\usepackage[english]{babel}
\usepackage{amsfonts}
\usepackage{amsmath}
\usepackage{array}
\usepackage{amsthm}
\usepackage{amssymb}
\usepackage{graphicx}
\usepackage{subfigure}
\usepackage{braket}
\usepackage{eucal}
\usepackage{verbatim}
\usepackage[table]{xcolor}
\usepackage{caption}
\usepackage{cite}
\usepackage{textcomp}
\usepackage{tikz}
\usetikzlibrary{positioning,arrows}
\usetikzlibrary{decorations.pathmorphing}
\usetikzlibrary{decorations.markings}
\usetikzlibrary{calc,decorations.markings}
\usetikzlibrary{arrows,shapes}
\usetikzlibrary{matrix,arrows}
\usepackage{pgfplots}
\usepackage{xparse}
\definecolor{jade}{HTML}{00A86B}
\newcommand{\be}{\begin{eqnarray}}
\newcommand{\ee}{\end{eqnarray}}

\renewcommand{\d}{\mbox{${\rm d}$}} 

\newcommand{\gn}{G_{\rm N}}

\title{\bf Accretion onto BNG stars}
\author{Florin Manea$^{a}$\thanks{E-mail: florin.manea1@s.unibuc.ro}
$\ $
and
Octavian Micu$^b$\thanks{E-mail: octavian.micu@spacescience.ro}
\\
\\
$^a${\em Faculty of Mathematics and Computer Science, University of Bucharest}
\\
{\em Iuliu Maniu Boulevard 15, Bucharest}
\\
\\
{\em $^b$Institute of Space Science, Bucharest, Romania}
\\
{\em P.O. Box MG-23, RO-077125 Bucharest-Magurele, Romania}
}
\begin{document}
\maketitle	
\begin{abstract}
We analyse the process of steady, spherically symmetric accretion onto BNG stars that are at rest with respect to a surrounding polytropic gas cloud which is at rest at infinity. We have shown that a transonic solution exists for the entire range of values spanned by the polytropic index of the gas, provided that the sound speed at infinity is below the corresponding critical value. We have shown that the accretion rate is independent of the stellar compactness and that, similarly to Bondi accretion, it decreases with increasing sound speed at infinity and polytropic index.

\end{abstract}
\section{Introduction}
\label{S:action}
\setcounter{equation}{0}

The strong-field regime of gravity is slowly becoming increasingly accessible to experiment, yet our understanding of the nature of the extremely compact objects that source these fields remains very limited. One of the toughest open questions to answer is whether the space-time singularities predicted by general relativity (GR) inside black holes really exist in nature, or if they just signal that the validity of the theory breaks down in the strong-field regime. The latter possibility led the community to the search for alternative solutions that admit regular interior solutions, potentially by incorporating quantum effects into the classical theory.

Bootstrapped Newtonian gravity (BNG) was introduced in Refs.~\cite{BootN, Casadio:2019cux} and further developed in a series of subsequent works \cite{Casadio:2019pli, Casadio:2020djs, Casadio:2020kbc, Casadio:2022pme, Casadio:2022gbv, Casadio:2025wwb} as a framework designed for investigating this question. There are multiple possible approaches for modifying GR, and BNG takes a bottom-up route: it starts from Newtonian gravity and adds non-linear interaction terms whose functional forms mirror those appearing at leading orders in the weak-field expansion of GR, treating all such terms on equal footing from the onset. 
The contribution of each additional term can be tuned through the dimensionless coupling constants, which makes the theory flexible and analytically tractable. 

Although extending a theory developed from the terms appearing in the weak-field regime into strong-fields may seem questionable, the approach has led to some very interesting results. We have shown that BNG admits interior solutions for extremely compact stars that remain regular even at compactness values for which GR predicts the necessary formation of a singularity. Equally notable is the absence of a Buchdahl limit~\cite{Buchdahl:1959zz} in BNG theory which means that, unlike in GR, there is no maximum compactness beyond which pressure necessarily fails to support the star in equilibrium. In other words BNG stars can in principle remain in equilibrium at arbitrarily large compactness values.

Another feature of the BNG theory is the existence of two different notions of mass. The ADM-like mass $M_\star$ represents the quantity that characterizes the gravitational field as seen by a distant observer, while the proper mass $M_{0,\star}$ is the volume integral of the (rest-mass) density. The two quantities are generally different, which is a direct consequence of the non-linear nature of the theory and its departure from Newtonian gravity. 

An important next step in the development of the formalism is to investigate the accretion process by BNG stars. We investigate accretion onto BNG stars of ADM-like mass $M_\star$ and compactness $X$, which are assumed to be at rest within a gas cloud described by the polytropic equation of state which, in turn, is at rest at infinity where it is characterised by a uniform density $\rho_\infty$ and a pressure $p_\infty$. The accretion process is spherically symmetric, the gas is in a steady motion, and its mass is assumed to be negligible, such that its contribution to the gravitational field can be neglected so as not to alter the exterior BNG gravitational potential. The same consideration applies to the increase in the star's mass. 
Spherical accretion was first studied by Bondi in the Newtonian framework~\cite{Bondi:1952ni}, and then extended to the relativistic regime for the accretion of polytropic gas onto a Schwarzschild black hole~\cite{Michel:1972oeq}. It was also generalised to modified theories of gravity, such as $f(R)$ gravity~\cite{Ahmed:2015tyi}, while spherically symmetric accretion onto alternative compact objects like boson stars has been investigated numerically \cite{Meliani:2016rfe}.

\section{BNG Stars}

The structure of a compact object in BNG is determined by a system of coupled differential equations: an Euler-Lagrange equation for the gravitational potential and a conservation equation which is responsible for the equilibrium between the gravitational pull and the pressure support. 
The Euler-Lagrange equation is obtained from the total Lagrangian 
\be 
L[V]
&\!\!=\!\!&
L_{\rm N}[V]
-4\,\pi
\int_0^\infty
r^2\,\d r
\left[
q_V\,\mathcal{J}_V\,V
+
q_p\,\mathcal{J}_p\,V
+
q_\rho\, \mathcal{J}_\rho \left(\rho+q_p\,\mathcal{J}_p\right)
\right]
\nonumber
\\
&\!\!=\!\!&
-4\,\pi
\int_0^\infty
r^2\,\d r
\left[
\frac{\left(V'\right)^2}{8\,\pi\,\gn}
\left(1-4\,q_V\, V\right)
+\left(\rho+\,q_p\,p\right)
V
\left(1-2\,q_\rho\, V\right)
\right]
\ ,
\label{LagrV}
\ee
in terms of the potential $V=V(r)$ (for static spherically symmetric objects)
and its derivatives with respect to the radial coordinate $r$ denoted by primes.
The contribution of the additional terms can be tuned through the coupling constants $q_V$, $q_p$ and $q_\rho$. When these are simultaneously taken to zero, all additional terms vanish and the Lagrangian reduces to the well-known Newtonian Lagrangian for a spherically symmetric object with a matter density $\rho=\rho(r)$ that generates a Newtonian gravitational potential $V_{\rm N}$. 
In general, in the previous investigations, all additional terms were assumed to contribute maximally (i.e., the coupling constants were set to their largest possible value), and the same is considered here $q_\rho = q_p = q_V \equiv 1$. However, since the accretion process depends only on the gravitational potential outside the source, which is determined only by the coupling $q_V$, the present analysis remains valid regardless of the values of $q_\rho$ and $q_p$.
The additional terms in the Lagrangian represent the following: the term proportional to $q_V$ is the gravitational self-coupling sourced by the gravitational energy $U_{\rm N}$ per unit volume
\be
\mathcal{J}_V
\simeq
\frac{\d U_{\rm N}}{\d \mathcal{V}} 
=
-\frac{\left[ V'(r) \right]^2}{2\,\pi\,\gn}
\ ,
\label{JV}
\ee
the contribution due to the pressure
\be
\mathcal{J}_p
\simeq
-\frac{\d U_p}{\d \mathcal{V}} 
=
p
\ ,
\label{JP}
\ee
which was introduced in Ref.~\cite{Casadio:2019cux}, in order to account for the effects of pressure which in the high compactness regime cannot be neglected, and a higher-order term
\be
\mathcal{J}_\rho
=
-2\,V^2
\ ,
\ee
which couples with the matter source. One of the effects of the (non-negligible) pressure term is to replace $\rho \to \rho+q_p\,p$ in the Lagrangian.

With the coupling constants set to one, the Euler-Lagrange equation for the potential resulting from Eq.~\eqref{LagrV} becomes
\be
\triangle V
=
4\,\pi\,\gn\left(\rho+p\right)
+
\frac{2 \left(V'\right)^2}
{1-4\,V} \ ,
\label{EOMV}
\ee
with $\triangle$ representing the Laplacian operator in spherical coordinates.

The GR conservation equation is given by the Tolman-Oppenheimer-Volkoff (TOV) equation~\cite{Tolman:1939jz,Oppenheimer:1939ne},
which is derived from the Einstein equations without the need of any additional assumptions.
The BNG equation of conservation, which expresses the equilibrium between the pressure gradient and the gravitational pull, must include the contribution of the pressure $p$ which is not negligible~\cite{Casadio:2019cux}, and with $q_p=1$ we have
\be
p'
=
-\left( \rho + p\right) V'
\ .
\label{eqP}
\ee

Depending on the type of star, an additional constraint is imposed by the equation of state which relates the pressure and density profiles, as in the case of stars characterised by the polytropic equation of state.

As demonstrated in our previous work, when solving the above system of differential equations, the non-linear nature of the theory generally prevents analytic solutions, therefore, numerical methods and approximations must usually be employed.
Both the interior and exterior solutions of the Euler-Lagrange equation have been investigated in the references cited in Section~\ref{S:action}. While the interior solutions change depending on the equation of state, resulting in various possible pressure and density profiles that lead to different interior gravitational potentials, the exterior solution for the potential is unique, given that in vacuum Eq.~\eqref{EOMV} reduces to 
\be
\triangle V
=
\frac{2 \left(V'\right)^2}
{1-4\,V} 
\ .
\label{EOMVout}
\ee

The gravitational potential is obtained separately inside and outside the source, while also ensuring continuity and smoothness at the boundary. In the outer vacuum Eq.~\eqref{EOMV} simplifies due to the density and pressure being null, which leads to the solution
\be
V
\equiv
V_{\rm out}
=
\frac{1}{4}
\left[
1-\left(1+\frac{6\,X}{\xi}\right)^{2/3}
\right]
\ ,
\label{sol0}
\ee
where $\xi = r/R$ represents the dimensionless radial coordinate. The two integration constants were fixed by requiring the solution to have the correct Newtonian behaviour for an object of mass $M_\star$ at a large distance from the source. It should be noted that the expansion of the above expression for large $r$ leads to the expected post-Newtonian term of the order $\gn^2$ as well.
The dimensionless parameter $X$ was introduced to define the compactness of the configuration in terms of the star radius $R$ and its ADM-like mass $M_\star$ as
\be
X \equiv \frac{\gn M_\star}{R} \ .
\ee

The ADM-like mass represents the mass parameter that characterises the object's (outer) gravitational potential, as inferred by a distant observer, for instance from analysing orbital motion. 
As stated in Section \ref{S:action}, in addition to the ADM-like mass, BNG stars are also characterized by their proper mass $M_{0, \star}$, defined as the volume integral of the matter density. 
\be
M_{0, \star}=4\, \pi \, R^3\int_0^1 \rho(\xi)\, \xi^2 \d \xi \ .
\label{M_0_def}
\ee
These two quantities were shown to be different \cite{Casadio:2019pli}. 

To determine the interior pressure and density profiles, along with the gravitational potential, one must solve the system of equations consisting of the Euler-Lagrange equation, the equation of state for the interior, and the continuity equation. The solutions are also subject to three boundary conditions: continuity of the potential and its first derivative at the stellar surface, where they are matched to the exterior solution given by Eq.~\eqref{sol0}, and the vanishing of the first derivative of the potential at the centre.
However, since our focus is on accretion onto BNG stars and the exterior gravitational potential depends only on their compactness $X$, we will not discuss their interior pressure and density profiles in detail.

\section{Accretion onto BNG stars}
\setcounter{equation}{0}

We recall that we consider the process of steady, spherically symmetric accretion onto a BNG star of ADM-like mass $M_\star$ and compactness $X$, at rest with respect to the surrounding gas at infinity, where the density and pressure approach the uniform values $\rho_\infty$, respectively $p_\infty$. The mass of the accreting gas is assumed to be negligible compared to that of the star, which allows us to neglect its gravitational contribution and treat the exterior BNG potential as fixed. The pressure and density of the accreting gas must also be sufficiently small compared to those inside the star so that the gas does not affect the background gravitational potential. Under this assumption, the proper mass of the accreting gas is equal to its ADM-like mass. The increase in the stellar mass due to accretion is also neglected.
Regarding the accreting gas, its pressure and density are related throughout the entire volume by the polytropic equation of state

\begin{equation}
p = \kappa\, \rho^\Gamma
\label{eq:polytropic_eos}
\end{equation}
where $1<\Gamma <\frac{5}{3}$ represents the polytropic index, with $\kappa$ a constant of proportionality that depends on the composition and entropy of the gas.

The (constant) rate of accretion at a certain radial coordinate $\xi\geq 1$ is encapsulated in the continuity equation 
\be
\dot{M}_{0} = 4\pi R^2\,\xi^2 \rho \, v \equiv {\rm const}
\ ,
\label{eq:continuity}
\ee
where the dot denotes differentiation with respect to time, $M_0$ denotes the proper mass of the accreted gas, and $v= \dot{\xi}$ represents the inward dimensionless radial velocity. Note that the speed of sound $c_s$ is also dimensionless throughout the paper. 

Following Bondi's treatment \cite{Bondi:1952ni}, we write the Bernoulli equation by anchoring the enthalpy term relative to a reference point at infinity, where the gas is assumed to be at rest and the gravitational potential vanishes. Outside the source, the BNG gravitational potential is given by the expression in Eq.~\eqref{sol0}, and the Bernoulli equation becomes

\be
\frac{v^2}{2} + \int_ {p_\infty}^p\,\frac{\d p}{\rho}+ V = 0
\label{eq:Bernoulli}
\ .
\ee
After performing the integral and replacing the pressure to density ratios, both at infinity and at $\xi$,  with the corresponding local speed of sound $c_s$ in the environment surrounding the BNG object using the expression
\be
c_s^2 
= 
\left( \frac{\partial p}{\partial \rho} \right)_s = \Gamma\, \frac{p}{\rho}
\equiv \Gamma\, \kappa\, \rho^{\Gamma-1}
\ ,
\label{eq:speed_of_sound}
\ee
we can write
\be
\frac{v^2}{2} + \frac{c_s^2}{\Gamma-1}  + V = \frac{c_{\infty}^2}{\Gamma-1}
\ ,
\label{Bern_Eq_gen}
\ee
where 
\be
c_{\infty}^2
= \Gamma\, \frac{p_\infty}{\rho_\infty}\  ,
\label{eq:speed_of_sound_infty}
\ee
represents the speed of sound at infinity. 

\subsection{Sonic radius and critical points}

A relevant quantity in the accretion process is the BNG sonic radius $\xi_s$, which represents the radius at which the fluid
flow speed equals the local speed of sound, transitioning from subsonic to supersonic~\cite{Bondi:1952ni}. 
After substituting Eq.~\eqref{eq:speed_of_sound} in Eq.~\eqref{Bern_Eq_gen} and differentiating the latter we have
\be
v\, v' + c_s^2\, \frac{\rho'}{\rho} + V' = 0 \ ,
\label{eq:BernoulliSubstituted}
\ee
where the prime denotes a derivative with respect to the dimensionless radial coordinate $\xi$. Differentiating the equation of continuity~\eqref{eq:continuity} yields
\be
\frac{\rho'}{\rho} = - \frac{2}{\xi} - \frac{v'}{v} \ .
\ee
After substituting this result in Eq.~\eqref{eq:BernoulliSubstituted} one finds 
\be
v' \,\frac{c_s^2-v^2}{v}=V' - \frac{2\, c_s^2}{\xi} \ .
\label{eq:grad_v}
\ee
When the radial velocity $v$ equals the local speed of sound $c_s$, the equation can only be satisfied if both sides vanish, leading to the following condition at the sonic radius:
\be
v_s^2=c_{s,s}^2\ ,
\qquad
\xi_s V'(\xi_s)=2c_{s,s}^2 \ ,
\label{sonic_pt}
\ee
where we labeled the speed of sound at the sonic radius by $c_{s,s}$. 

After substituting the above expression for the speed of sound, as well as the exterior BNG potential and its first derivative in Eq.~\eqref{Bern_Eq_gen}, the equation that determines the sonic radius $\xi_s$ becomes
\be
& &
  \frac{(\Gamma+1)\,\xi_s V'(\xi_s)}{4\,(\Gamma-1)}
  +V(\xi_s)-\frac{c_\infty^2}{\Gamma-1} \nonumber
\\
& &=\frac{1}{4}\left[
1-\left(1+\frac{6X}{\xi_s}\right)^{2/3}
+\frac{X(1+\Gamma)}
{(\Gamma-1)\xi_s(1+6X/\xi_s)^{1/3}}
\right]
-
 \frac{c_{\infty}^2}{\Gamma-1}  \nonumber
\\
& &\equiv
  F(\xi_s) =0 
 \ .
 \label{eq:F}
\ee 

Note that the sonic radius only depends on the speed of sound at infinity $c_\infty$, the polytropic index of the surrounding environment $\Gamma$ and the stellar compactness $X$. Thus, stars with different ADM-like masses but the same compactness have the same value of $\xi_s$.
Upon careful analysis, the equation above can have up to three real solutions, of which only the positive ones are of interest. To be even more exact, the only relevant physical solutions are for $\xi_s\geq1$, where the equality corresponds to the limiting case in which the flow becomes sonic at the stellar surface. We can now analyse the existence of a sonic radius by investigating the number of positive (and larger than one) roots of the function $ F(\xi_s)$ with the help of its first and second derivatives. 

Setting the first derivative of $F$ to zero determines the extrema of the function
\be
 F(\xi_s)'
 &=&\frac{(5\Gamma-3)V'(\xi_s)+(\Gamma+1)\xi_s V''(\xi_s)}{4(\Gamma-1)}
 \nonumber \\
&=&\frac{
X\left[4X(5\Gamma -7)+(3\Gamma - 5)\xi_s\right]
}{
4(\Gamma-1)\xi_s^{5/3}(6X+\xi_s)^{4/3}
}
=
0
\ .
\ee

Given the constraints on $\xi_s$, the denominator is always positive, and the zeros of $F'$ are therefore obtained by setting its numerator to zero.
In the interval $1<\Gamma\leq7/5$, the function $F'$ has no zeros in the physical domain $\xi_s \geq 1$. 
Consequently, $F$ is strictly monotonic and it can admit at most one positive root. 
In the range $1<\Gamma<7/5$, $\lim_{\xi_s\to0}F(\xi_s)= + \infty$, while $\lim_{\xi_s\to\infty}F(\xi_s)=-c_{\infty}^2/(\Gamma-1)<0$. 
For $\Gamma=7/5$, while the limit at infinity remains the same, $\lim_{\xi_s\to0}F(\xi_s)=1/4-5\,c_{\infty}^2/2$ is positive for any speed of sound at infinity smaller than $c_{\infty}<1/\sqrt{10}\simeq0.316$, i.e., roughly a third of the speed of light. 
This places this particular case in the same regime as the rest of the interval, in the sense that the function is positive near the centre and negative at infinity. 
Therefore, a unique positive root always exists. However, this solution is physical only if it is located outside the object, at $\xi_s>1$, where the equations from which it is derived are valid.

For $7/5<\Gamma<5/3$, $F'$ has a single positive zero, and consequently $F$ has a single extremum. Depending on the parameters and on the value of $F$ at this extremum, the equation $F(\xi_s)=0$ can admit zero, one, or two distinct positive solutions.
The second derivative of $F$ is calculated to be

\be
 F(\xi_s)''
 =
\frac{
X\left[
20X^2(7-5\Gamma)
+4X(13-9\Gamma)\xi_s
+(5-3\Gamma)\xi_s^2
\right]
}{
2(\Gamma-1)\xi_s^{8/3}(6X+\xi_s)^{7/3}
}
\ ,
\ee
and can be used to determine whether the extremum is a local minimum or maximum. The location of the extremum is found by setting $F'(\xi_s)=0$, which yields

\be
\xi_m = -\frac{4 X(7-5\Gamma)}{(5-3\Gamma)}
\ .
\label{xi_m}
\ee 
By replacing this result into the expression for the second derivative of $F$, one obtains 
\be
 F(\xi_m)''
 =
 -
\frac{(5-3\Gamma)^4}
{64\,2^{2/3}X^2(\Gamma-1)(1+\Gamma)^{4/3}(5\Gamma - 7)^{5/3}}
\ ,
\ee
which in the interval $7/5 < \Gamma< 5/3$ is always negative, meaning that $\xi_m$ is a local maximum. From Eq.~\eqref{xi_m}, for $\Gamma=5/3$ the location of the extremum is pushed to infinity. 

\begin{figure}[t]
\centering
\includegraphics[width=10cm]{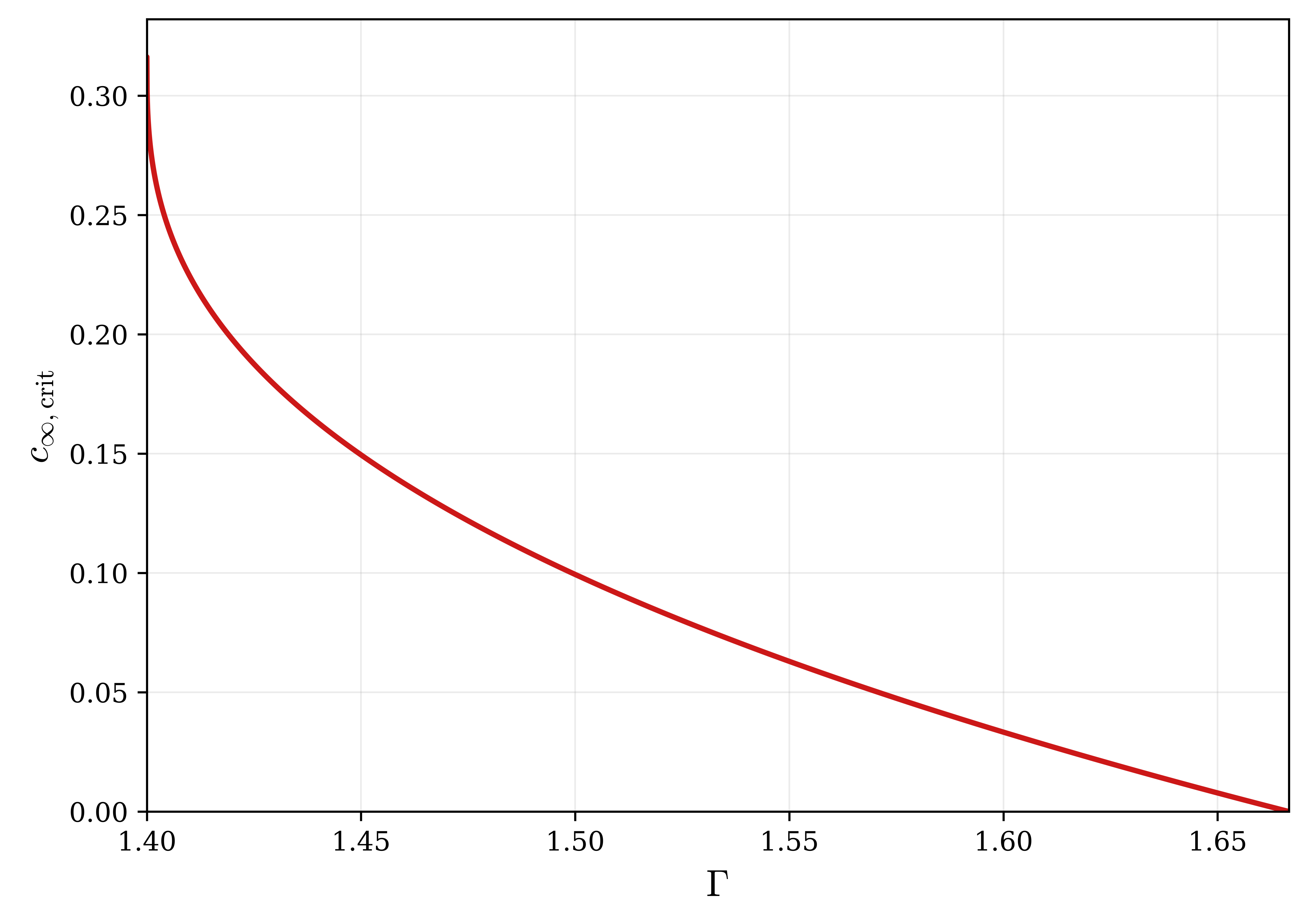}
\caption{$c_{\infty,\rm \rm crit}$ for polytropic index values in the range $7/5 < \Gamma< 5/3$.}
\label{p:c_crit_vs_gamma}
\end{figure}

Within this interval, the boundary separating the regimes with two positive roots and no positive roots is determined by the value $c_{\infty,\rm crit}(\Gamma)\equiv c_{\infty,\rm crit}$, corresponding to a single positive double root. 
To be more accurate, the critical condition $F(\xi_s)=F'(\xi_s)=0$ defines a critical curve in the $(\Gamma, c_{\infty}^2)$ parameter space. Although this relation is usually expressed as $\Gamma_{\rm crit}(c_{\infty})$, in this case it is more convenient to obtain a closed form as $c_{\infty,\rm crit}^2(\Gamma)$. Substituting the extremum location from Eq.~~\eqref{xi_m} into $F(\xi_s)=0$ and solving for $c_{\infty}^2$ yields, independently of $X$,
\be
c_{\infty,\rm crit}^2 = \frac{\Gamma-1}{4} - \frac{1}{16}\left[2(\Gamma+1)^2(5\Gamma-7)\right]^{1/3} \ .
\label{csinf_crit}
\ee
The critical curve is shown graphically in the plot in Fig.~\ref{p:c_crit_vs_gamma}. 
For $c_{\infty} < c_{\infty,\rm crit}$ $F$ admits two positive roots, for $c_{\infty} = c_{\infty,\rm crit}$ it admits a single positive double root, while for $c_{\infty} > c_{\infty,\rm crit}$ it admits none.

To determine which of the multiple roots of $F(\xi_s)=0$ corresponds to a physically realizable accretion flow, we examine the local topology of the flow near each critical point~\cite{Abramowicz:2011xu, Mandal:2007pu, Chaudhury:2006jf, Ray:2002sz}. This is done by analysing how the radial velocity of the flow changes as it approaches the star, since only trajectories that connect continuously to the boundary condition at infinity describe an admissible accretion flow. The expression from Eq.~\eqref{eq:grad_v} can also be written as
\be
v' = \frac{ v \left(\frac{2\,c_s^2}{\xi}- V' \right)} {v^2 - c_s^2}
\equiv
\frac{N(\xi,v)}{D(\xi,v)} \ .
\ee
At a critical point $\xi_s$, both the numerator $N(\xi,v)$ and the denominator $D(\xi,v)$ must vanish simultaneously, in agreement with Eq.~\eqref{sonic_pt}.

To investigate the nature of each critical point $(\xi_s, v_s)$, we treat $\xi$ and $v$ as independent variables of a two-dimensional dynamical system by introducing an auxiliary parameter $\tau$ and rewriting the equation governing the radial velocity as
\be
\frac{\d \xi}{\d\tau}
=
D(\xi,v)
\ ,
\qquad
\frac{\d v}{\d\tau}
=
N(\xi,v)
\ .
\ee

Next we linearize the system around a critical point $(\xi_s, v_s)$, introducing small perturbations $\delta\xi$ and $\delta v$. To first order, the resulting system takes the form
\be
\frac{d}{d\tau}
\begin{pmatrix}
\delta\xi\\
\delta v
\end{pmatrix}
=
J_s
\begin{pmatrix}
\delta\xi\\
\delta v
\end{pmatrix} \ ,
\ee
where $J_s$ is the Jacobian matrix evaluated at the critical point. 

In calculating the Jacobian matrix, the speed of sound $c_s$ must be treated as a function of $\xi$ and $v$, rather than a constant. 
The polytropic equation of state, together with the continuity equation for a constant rate of accretion, can be substituted in Eq.~\eqref{eq:speed_of_sound} to obtain
\be
c_s^2(\xi,v) 
=
\Gamma \kappa \left(\frac{\dot M_0}{4\pi R^2}\right)^{\Gamma-1}
\left(\xi^{2}\,v\right)^{1-\Gamma} 
\equiv
 \Gamma \kappa \mathcal{A}^{\Gamma-1}\left(\xi^{2}\,v\right)^{1-\Gamma} 
 \ ,
 \label{cq:cs2xiv}
\ee
where the constant $\mathcal{A}$ was defined to include only certain constant factors, for reasons that will become apparent later.

The partial derivatives of $c_s^2$ needed to evaluate the Jacobian matrix elements are calculated to be
\be
\frac{\partial c_s^2}{\partial v} = -\frac{(\Gamma-1)\,c_s^2}{v} \ , 
\ee
respectively
\be
\frac{\partial c_s^2}{\partial \xi} = -\frac{2(\Gamma-1)\,c_s^2}{\xi} \ .
\ee

Finally, the Jacobian matrix evaluated at $\xi_s$ is

\be
J_s =
\left(
\begin{array}{cc}
\frac{2\,(\Gamma-1)\,c_{s,s}^{2}}{\xi_s}
&
(\Gamma+1)\,v_s
\\[10pt]
-v_s\left[
\frac{2\,(2\,\Gamma-1)c_{s,s}^{2}}{\xi_s^{2}}
+V''(\xi_s)
\right]
&
-\frac{2\,(\Gamma-1)\,c_{s,s}^{2}}{\xi_s}
\end{array}
\right)
\ ,
\ee
where Eq.~\eqref{sonic_pt} was used to simplify the expressions of the $(J_s)_{12}$ and $(J_s)_{22}$ of the Jacobian matrix.  
Also, note that $\mathrm{tr} J_s=0$, a detail relevant for classifying the critical points. 

The eigenvalues of $J_s$ satisfy $\lambda_1\,\lambda_2=\det J_s$. Since the Jacobian matrix is traceless, the critical point is of saddle-type when $\det J_s<0$, corresponding to two real eigenvalues of opposite signs, and of centre-type when $\det J_s>0$, corresponding to two purely imaginary eigenvalues. Only a saddle-type critical point allows a transonic trajectory to pass smoothly through it, whereas trajectories near a centre-type point circulate around it without crossing the critical point~\cite{Abramowicz:2011xu, Mandal:2007pu, Chaudhury:2006jf, Ray:2002sz}. The determinant of the Jacobian simplifies to

\be
\det J_s 
&=&
c_{s,s}^2
\left[
(\Gamma+1)V''(\xi_s)
+(5\Gamma-3)\frac{V'(\xi_s)}{\xi_s}
\right] \nonumber\\
&=& \frac{4\,(\Gamma-1)\,c_{s,s}^2}{\xi_s} \,F'(\xi_s)
\ .
\ee
Since the prefactor is positive for $\Gamma>1$ and $\xi_s>0$, the sign of the Jacobian determinant is determined entirely by the sign of $F'(\xi_s)$.

We recall that in the interval $7/5<\Gamma<5/3$, for $c_{\infty}<c_{\infty,\mathrm{crit}}$, $F(\xi_s)=0$ has exactly two positive roots $\xi_1$ and $\xi_2$. The derivative $F'(\xi_s)$ vanishes at $\xi_m$, where $F(\xi_s)$ reaches a maximum, and where the derivative also changes sign. The roots satisfy the following relationship $\xi_1<\xi_m<\xi_2$. Given that $F'(\xi_1)>0$ for $0<\xi_1<\xi_m$, and $F'(\xi_2)<0$ for $\xi_2>\xi_m$ we have
\be
& &\det J_s(\xi_1)>0 \quad \Rightarrow \quad \xi_1 \ \text{is unphysical (of centre-type)} \ ,\\
& &\det J_s(\xi_2)<0 \quad \Rightarrow \quad \xi_2 \ \text{is physical (of saddle-type)} \ .
\ee
Therefore, whenever Eq.~\eqref{eq:F} has two positive roots, the inner root located at $\xi_1$ corresponds to a centre-type critical point, which does not admit a physical transonic solution, while the outer root located at $\xi_2$ corresponds to a saddle-type critical point. Depending on the particular values of the parameters that determine its value, when the latter solution also satisfies $\xi_2>1$, thus being located outside the object, it supports a physically admissible transonic accretion flow. 

At the critical parameter values for which the two roots merge, $F(\xi_s)=F'(\xi_s)=0$, the determinant of the Jacobian matrix vanishes and the resulting critical point is degenerate. Going back to the range $1<\Gamma \leq 7/5$, we have established that a single positive root exists. In this interval, we have shown that $F(\xi_s)$ is monotonically decreasing with $F'(\xi_s)<0$ throughout the entire positive range. Provided that $\xi_s>1$ the Jacobian determinant is negative, implying that this critical point is always of saddle-type.  
\begin{figure}[t]
\centering
\includegraphics[width=7cm]{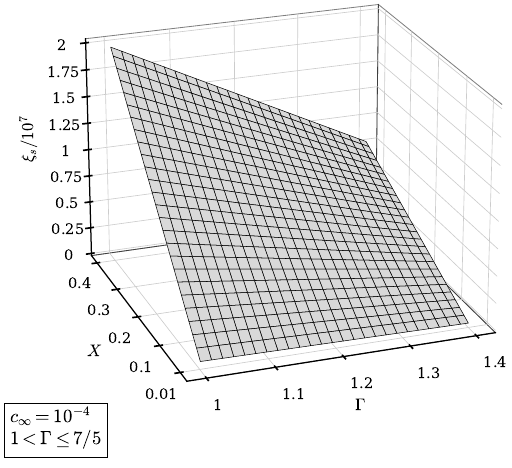}
\includegraphics[width=7cm]{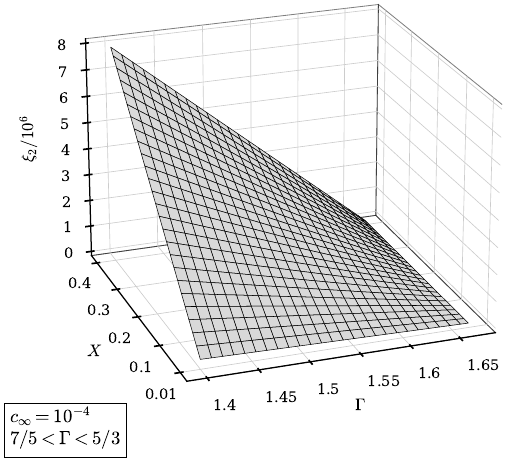}
\includegraphics[width=7cm]{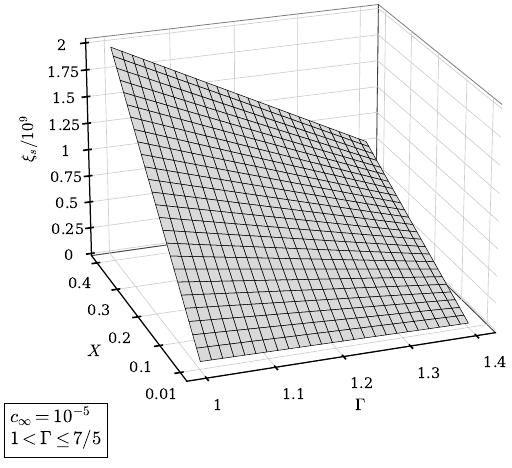}
\includegraphics[width=7cm]{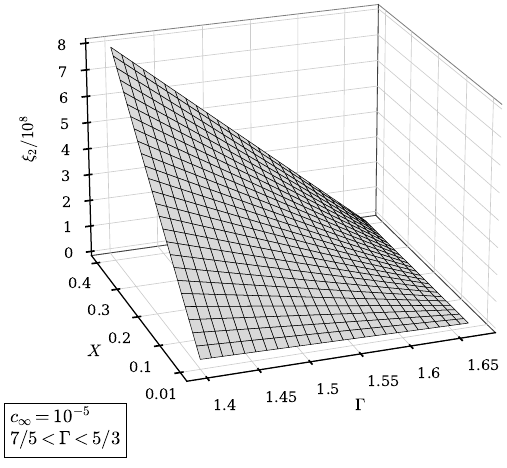}
\caption{The sonic radius $\xi_s$ (given by the outer root $\xi_2$ in the upper range of $\Gamma$) as a function of the compactness $X$ and the polytropic index $\Gamma$ for two values of the sound speed at infinity $c_\infty$.
}
\label{fig:xi_s}
\end{figure}

In the panels of Fig.~\ref{fig:xi_s} we represent the sonic radius as a function of the compactness $X$ and polytropic index $\Gamma$. 
For the numerical simulations, in the remainder of the paper we use two asymptotic values for the sound speed of the polytropic gas at infinity: $c_\infty=10^{-4}$ corresponding to approximately $30\,\mathrm{km\,s^{-1}}$, respectively $c_\infty=10^{-5}$ corresponding to about $3\,\mathrm{km\,s^{-1}}$. These values are representative of two different thermal conditions in the interstellar medium, both with sound speeds well within the non-relativistic regime~\cite{Toropina:2011kz}. The left panels show the unique positive root for the lower range of $\Gamma$, while the right panels show the outer root $\xi_2$, corresponding to a saddle-type critical point that admits transonic flow, and thus represents $\xi_s$ for this range of $\Gamma$. Throughout the entire parameter ranges considered for $X$ and $\Gamma$ the condition for $\xi_s >1$ is also satisfied. The values of $\xi_s$ and $\xi_2$ decrease below one only for compactness values several orders of magnitude smaller.

What remains to be verified is whether the transonic solution can be extended continuously all the way to the surface of the star. This can be investigated using the continuity equation for a constant rate of accretion, along with the Bernoulli equation. With the constant $\mathcal{A}$ defined in Eq.~\eqref{cq:cs2xiv} the Bernoulli equation becomes
\be
H(\xi, v) = \frac{v^2}{2} + \frac{ \Gamma \kappa}{\Gamma-1}  \left(\frac{\mathcal{A}}{\xi^{2}\,v}\right)^{\Gamma-1} + V(\xi) - \frac{c_{\infty}^2}{\Gamma-1}\equiv 0
\ .
\label{Bern_Eq_in}
\ee
Although the origin of both functions $F(\xi_s)$ and $H(\xi, v)$ is the Bernoulli equation, we adopted different notations to avoid confusion due to the different particular cases they represent.

For fixed values of all other parameters, the function $H(\xi, v)=0$ represents a curve in the $(\xi, v)$ plane. 
The more cumbersome approach to determining whether the transonic solution extends continuously from $\xi_s$ to the stellar surface located at $\xi=1$ is to solve explicitly for $v$ throughout this interval. However, there is a simpler way to determine whether a supersonic solution exists at every point in this interval. In order to do this, we assume an arbitrary but fixed value of $\xi$ in this range and investigate what conditions are needed to be fulfilled  for such a solution to exist. 
We notice that $\lim_{v\to0^+}H(\xi,v)=\infty$ due to the divergence of the second, enthalpy term dependent on the velocity, while $\lim_{v\to \infty}H(\xi,v)=\infty$ due to the divergence of the first, kinetic term. Therefore, the function $H(\xi,v)$ must have at least one minimum for some intermediate value of the velocity $v>0$. 
This value can be determined by taking the first derivative of Eq.~\eqref{Bern_Eq_in} with respect to $v$, and setting it to zero. After some algebraic manipulation and substitutions, the equation simplifies to
\be
\frac{\partial H(\xi, v)}{\partial v} = \frac{v^2-c_s^2}{v} = 0 
\ ,
\label{Bern_Eq_in_partial}
\ee
 which means that the function $H(\xi,v)$ has one positive minimum for $v=c_s$. 
 Considering the two limits of the function discussed in the paragraph above and the existence of a single minimum, we conclude that the equation $H(\xi,v)=0$ admits two (distinct) positive roots when $H(\xi,c_s)<0$, a single degenerate root when $H(\xi,c_s)=0$, and no positive roots when $H(\xi,c_s)>0$. Whenever two positive roots exist, the smaller one always corresponds to subsonic flow, while the larger corresponds to supersonic flow. 
 
The minimum of $H(\xi,v)$ in terms of an arbitrary radius $1<\xi <\xi_s$ where the corresponding local speed of sound is $c_{s,0}$ is 

\be
H_{\min}(\xi, c_{s,  0}) = \frac{\Gamma+1}{2\left( \Gamma-1\right)}c_{s,0}^2 + V(\xi) - \frac{c_{\infty}^2}{\Gamma-1} 
\ .
\label{Bern_Eq_in_a}
\ee

The expression for $c_{s,  0}$ is obtained from Eq.~\eqref{cq:cs2xiv} by setting the inward radial velocity equal to the sound speed and solving for the latter
\be
c_{s,  0}^2
=
\left(\Gamma \kappa\right)^{\frac{2}{\Gamma+1}} \left(\frac{\mathcal{A}}{\xi^{2}}\right)^{\frac{2(\Gamma-1)}{\Gamma+1}} 
\ .
\ee

By substituting back into Eq.~\eqref{Bern_Eq_in} results an expression for the minimum of the function $H_{\min}(\xi)$ that depends only on the dimensionless radial coordinate $\xi$. The derivative of this resulting function with respect to $\xi$ is 
\be
\frac{\d H_{\min}(\xi)}{\d \xi} = V'(\xi)-\frac{2\,c_{s, 0}^2}{\xi}\ ,
\ee
meaning that the extremum of $H_{\min}(\xi)$ satisfies the constraint $\xi\,V'(\xi)=2\,c_{s, 0}^2$, which is precisely the condition from Eq.~\eqref{sonic_pt}. At the exterior saddle point we have simultaneously $H_{\min}(\xi_2)=0$ and $H'_{\min}(\xi_2)=0$.

The only remaining task is to check whether the function $H_{\min}(\xi) \leq 0$ throughout the interval $1<\xi<\xi_s$. 
It should be emphasized that, in general, the value of the constant $\mathcal{A}$ calculated at the centre-type critical point $\xi_1$ is different from the value evaluated at the saddle-type critical point $\xi_2$, the latter being the relevant solution for the physical accretion flow.

\begin{figure}[t]
\centering
\includegraphics[width=5.1cm]{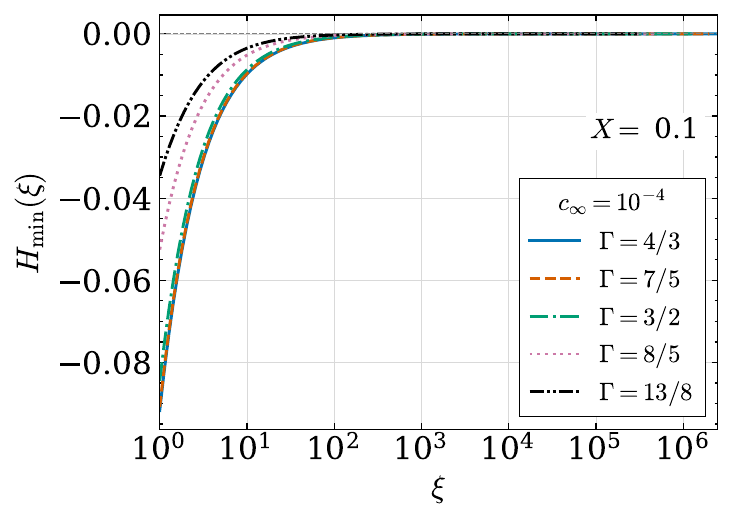}
\includegraphics[width=5.1cm]{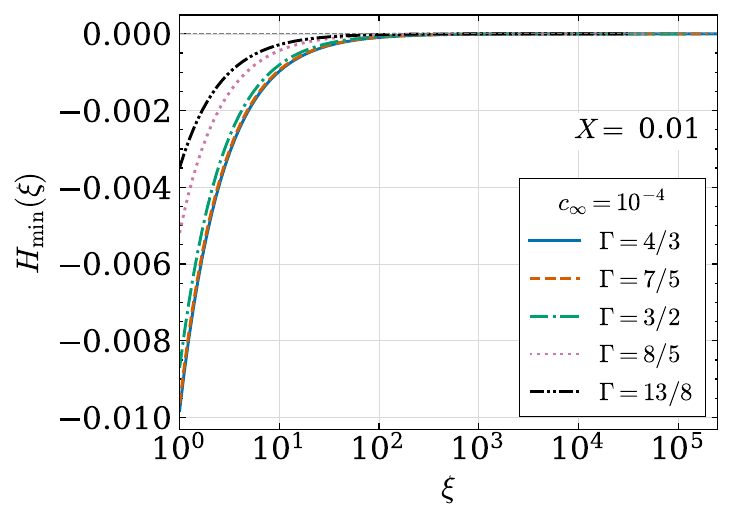}
\includegraphics[width=5.1cm]{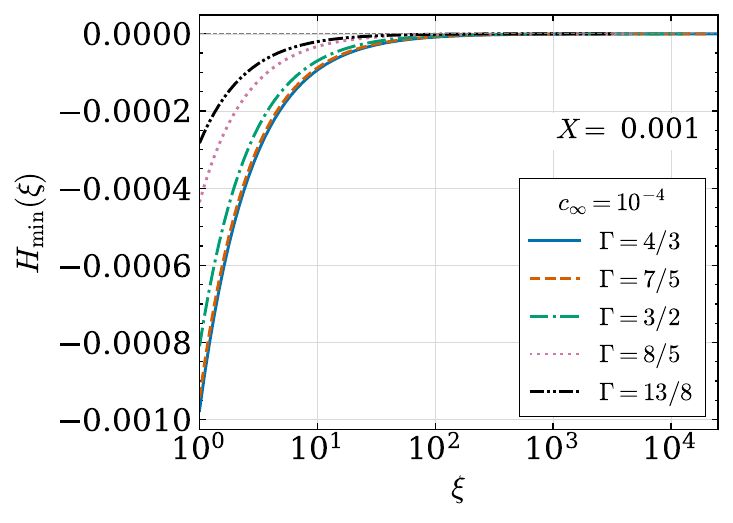}
\includegraphics[width=5.1cm]{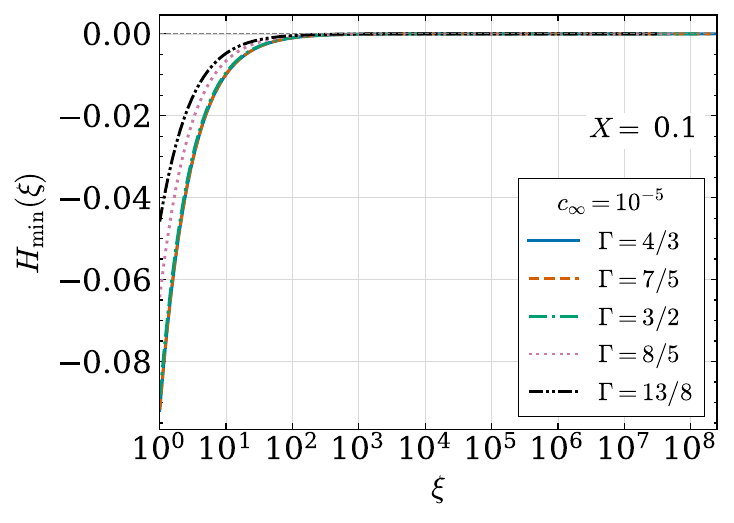}
\includegraphics[width=5.1cm]{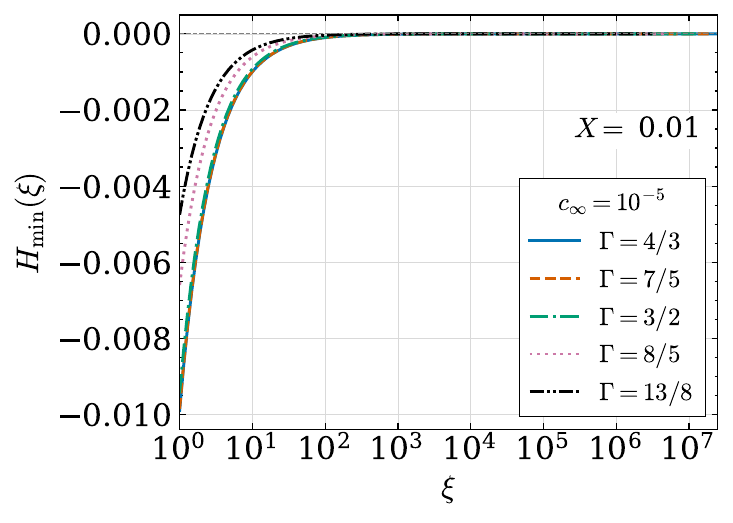}
\includegraphics[width=5.1cm]{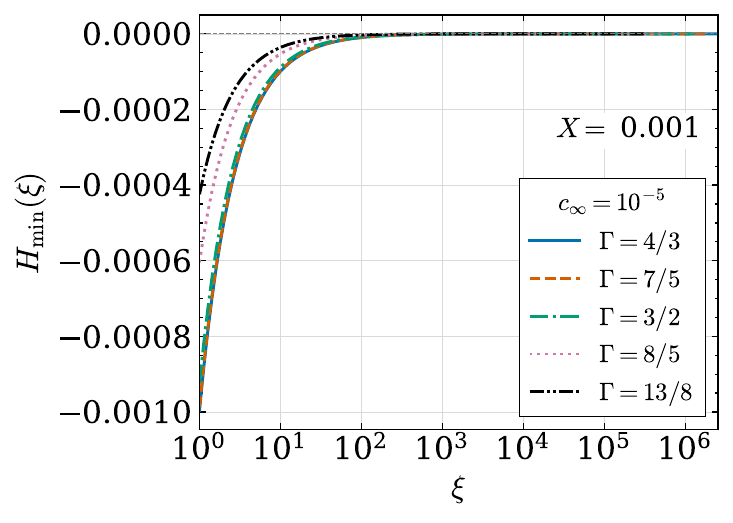}
\caption{The function $H_{\min}(\xi)$ for three values of the compactness $X$, different values of the polytropic index $\Gamma$, and two values of the sound speed at infinity $c_\infty$.}
\label{Hmin}
\end{figure}

To calculate $\mathcal{A}=\xi_s^2\,\rho_s\,c_{s,s}$ the ratio of the speeds of sound at the sonic point and at infinity is used to express the density at the sonic point as 
\be
\rho_s=\rho_\infty \left(\frac{c_{s,s}^2}{c_\infty^2}\right)^{\frac{1}{\Gamma-1}}
=
\rho_\infty \left[\frac{\xi_s\,V'(\xi_s)}{2\,c_\infty^2}\right]^{\frac{1}{\Gamma-1}}
\ .
\ee
With this and the expression of the speed of sound we finally have 
\be
\mathcal{A}
=
2^{-\frac{\Gamma+1}{2(\Gamma-1)}}
\rho_\infty\,
c_\infty^{-\frac{2}{\Gamma-1}}\,
X^{\frac{\Gamma+1}{2(\Gamma-1)}}\,
\xi_s^{\frac{3\Gamma-5}{2(\Gamma-1)}}\,
\left(1+\frac{6X}{\xi_s}\right)^{
-\frac{\Gamma+1}{6(\Gamma-1)}
}
\ ,
\label{def:A}
\ee
with $\xi_s$ determined using Eq.~\eqref{eq:F}. 
After substituting in all factors (except the expression for $\mathcal{A}$ which is somewhat cumbersome), we obtain 
\be
H_{\min}(\xi) 
 = \frac{\Gamma+1}{2\left( \Gamma-1\right)}\,c_\infty^{\frac{4}{\Gamma+1}} \left(\frac{\mathcal{A}}{\rho_\infty\,\xi^{2}}\right)^{\frac{2(\Gamma-1)}{\Gamma+1}} + \frac{1}{4}
\left[
1-\left(1+\frac{6\,X}{\xi}\right)^{2/3}
\right]
 - \frac{c_{\infty}^2}{\Gamma-1} 
\ .
\label{eq:Hmin}
\ee
We observe that, all other parameters such as the characteristics of the accreted gas being equal, stars with different ADM-like masses but the same compactness have the same $H_{\min}(\xi)$. Also, note that $H_{\min}(\xi)$ does not depend on the density of the gas at infinity, since the factor $\rho_\infty$ is canceled by that from the expression for $\mathcal{A}$. Figure~\ref{Hmin} shows plots of $H_{\min}(\xi)$ over the interval $\xi\in(1,\xi_s)$ for three values of the compactness $X$ and several values of $\Gamma$. Again, the same two values of the asymptotic sound speed of the polytropic gas, $c_\infty=10^{-4}$ and $c_\infty=10^{-5}$, were used. In all the cases, the function $H_{\min}(\xi)$ remains negative throughout the interval, thus confirming that a transonic branch that extends to the surface of the star exists in each case.

\subsection{Mass accretion rate}

For a steady, spherically symmetric flow, the conserved rest-mass accretion rate is given by Eq.~\eqref{eq:continuity}. 
At the sonic point, $\xi=\xi_s$, the flow speed equals the local sound speed, $v_s=c_{s,s}$. Therefore, we can compute the accretion rate using
\be
\dot M_0 &=& 4\pi R^2  \mathcal{A}
\nonumber
\\
&=&
4\pi\left(\frac{G_{\rm N}M_\star}{X}\right)^2 \mathcal{A}
\ ,
\ee
after substituting the radius $R$ using the expression for the compactness. The expression for $\mathcal{A}$ is given in Eq.~\eqref{def:A}. 
So far, we have worked in units where $c=1$. To calculate the physical mass accretion rate, we must restore the appropriate factors of the speed of light
\begin{equation}
\dot M_0^{\rm phys} =
\frac{4\pi G_{\rm N}^2\,M_\star^2}{X^2\,c^3}
\mathcal{A}
\ ,
\label{Mdotphys}
\end{equation}
where $\mathcal{A}$ has the dimensions of mass density that result from the factor $\rho_\infty$, for which we take the value $\rho_\infty=10^{-24}\, \mathrm{g\,cm^{-3}}$. 

The expressions above suggest that the accretion rate may depend on the compactness $X$, and we shall investigate this aspect in more detail. By making the substitution $Y=X/\xi_s$ in Eq.~\eqref{eq:F}, we obtain
\be
\frac{1}{4}\left[
1-\left(1+6\,Y\right)^{2/3}
+\frac{Y(1+\Gamma)}
{(\Gamma-1)(1+6\,Y)^{1/3}}
\right]
-
 \frac{c_{\infty}^2}{\Gamma-1} 
 = 0 
 \ ,
\ee
equation that no longer depends explicitly on $\xi_s$ or $X$. The variable $Y$ represents the ratio of the gravitational length scale $\gn\,M_\star$ to the physical sonic radius $r_s$. The solutions of the equation above are functions of $\Gamma$ and $c_\infty$ only, $Y\equiv Y(\Gamma,c_\infty)$. Therefore, the dependence of $\xi_s$ on the compactness is linear,
\be
\xi_s=\frac{X}{Y(\Gamma, c_{\infty})}
\ .
\label{eq:Y}
\ee
By substituting this expression in Eq.~\eqref{def:A}, it can be rewritten as
\be
\mathcal{A}
=
X^2\,
2^{-\frac{\Gamma+1}{2(\Gamma-1)}}
\rho_\infty\,
c_\infty^{-\frac{2}{\Gamma-1}}\,
Y(\Gamma, c_{\infty})^{-\frac{3\Gamma-5}{2(\Gamma-1)}}\,
\left[1+6\,Y(\Gamma, c_{\infty})\right]^{
-\frac{\Gamma+1}{6(\Gamma-1)}
}
\ .
\label{def:A-Y}
\ee
When substituting this expression into Eq.~\eqref{Mdotphys}, all factors of $X$ cancel out, proving that the accretion rate is independent of the compactness and that, for certain properties of the accreting gas, it depends only on the ADM-like mass of the star $M_\star$.

\begin{figure}[t]
\centering
\includegraphics[width=7cm]{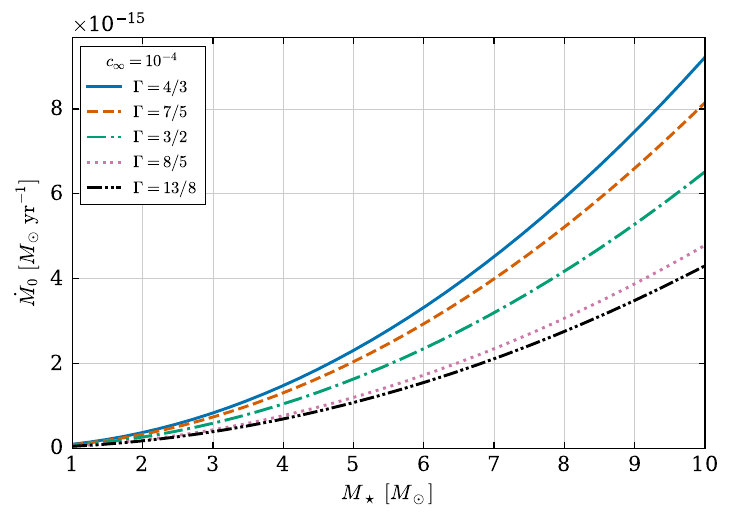}
\includegraphics[width=7cm]{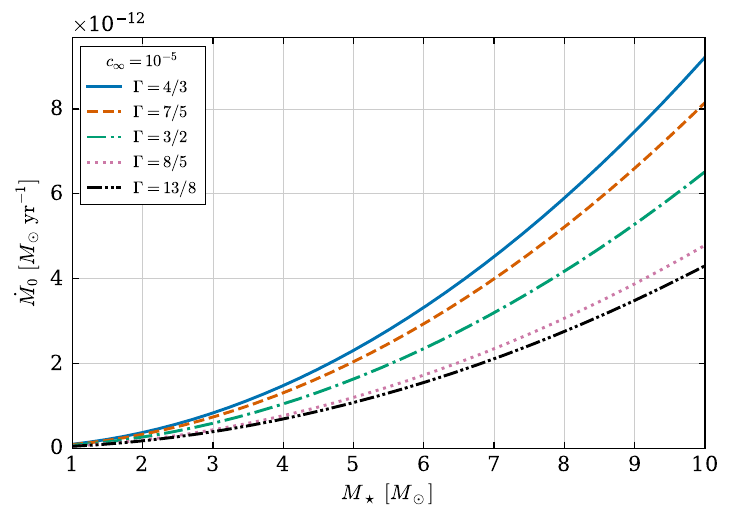}
\caption{The proper mass accretion rate $\dot M_0$ as a function of the ADM-like mass $M_\star$ of the accreting star for different values of 
$\Gamma$. Left panel: results for $c_\infty=10^{-4}$. Right panel: results for $c_\infty=10^{-5}$.}
\label{Mdot_Msun}
\end{figure}

In Fig.~\ref{Mdot_Msun} we plot the accretion rate $\dot M_0$ as a function of the ADM-like mass $M_\star$ of the BNG star in units of the solar mass $M_\odot$ for two values of $c_\infty$ and several values of the polytropic index $\Gamma$. We observe that the accretion rate decreases with the increase of $\Gamma$, vertical cuts through the plots representing accretion rates for stars of the same mass but different 
$\Gamma$ values. We chose to show a range of one order of magnitude in $M_\star$, since the accretion rate scales simply as $M_\star^2$. 

Note that the $M_\star^2$ scaling is also characteristic of Bondi accretion. However, as can be seen from the Lagrangian from Eq.~\eqref{LagrV}, in the outer vacuum the only surviving additional term is that proportional to $q_V$, and the Newtonian Lagrangian is recovered in the limit $q_V \rightarrow 0$. Since $q_V$ was set to its maximum value of one at the outset of the derivation and cannot be tracked analytically through the calculations, obtaining the Bondi accretion limit is not feasible within the present framework.

The accretion timescale is calculated by taking the ratio of the stellar mass to the mass accretion rate $\tau \simeq M_{0, \star}/\dot{M}_0$. 
However, in our case, the quantities at our disposal are the ADM-like mass of the star and the accretion rate of the gas proper mass, under the assumptions discussed at the beginning of this section. In order to calculate the accretion timescale consistently, one would need either to convert the proper mass accretion rate into an ADM-like mass accretion rate or to determine the proper mass of the star from its ADM-like mass. For BNG stars, both conversions depend on the stellar interior density, pressure profiles, and compactness $X$. Such an analysis is beyond the scope of the present investigation, which only requires knowledge of the exterior BNG gravitational potential.

\section{Conclusions and Discussion}
In this paper, we have further developed the BNG model by investigating the accretion process. While most previous studies, with a few exceptions~\cite{Casadio:2021gdf}, have focused on the interior of BNG stars, the accretion process is governed by the exterior gravitational potential.

We have shown that when the equation for the sonic radius admits two positive solutions, only the outer one (corresponding to a saddle-type critical point) can yield a physically admissible transonic accretion solution. The inner solution (associated with a centre-type critical point) does not allow such a solution~\cite{Abramowicz:2011xu, Mandal:2007pu, Chaudhury:2006jf, Ray:2002sz}. Additionally, besides the existence of a saddle-type critical point the sonic radius must lie outside the stellar surface and the corresponding flow must satisfy the appropriate boundary conditions.
In summary, for $1<\Gamma\leq 7/5$, the sonic radius $\xi_s$ corresponds to a unique positive root, while for $7/5<\Gamma< 5/3$, it corresponds to the outer root when two positive roots exist, or the degenerate double root when the Jacobian determinant vanishes. Of course, the solutions in question must lie outside the object, i.e. at $\xi_s>1$. 

Similarly to the Bondi accretion case, the accretion rate decreases with the increase of the gas speed of sound at infinity. The accretion rate also decreases with the increase of the polytropic index $\Gamma$. These behaviors are both explained by better pressure support that counteracts gravitational collapse. We have also proved analytically that the accretion rate is independent of the stellar compactness, implying that stars with the same ADM-like mass $M_\star$ have identical accretion rates for given properties of the surrounding gas.

A natural extension of the present work would be to investigate the accretion flow between the sonic radius and the stellar surface, particularly what happens at the surface. Another possible direction would be to use the accretion rates obtained here to study the time evolution of BNG stars, including changes in their interior density and pressure profiles as matter is accreted. Both investigations would be based on existing or newly developed models of BNG stellar interiors. 
These developments should account for the relationship between the proper mass $M_{0, \star}$ and the ADM-like mass $M_\star$, as well as for the evolution of the interior density, pressure, and gravitational potential as accretion proceeds. Such investigations are beyond the scope of the present work.
\section*{Acknowledgments}
Octavian Micu is supported by the Romanian Ministry of Research, Innovation and Digitalization
under the Romanian National Core Program LAPLAS VII-contract n.~30N/2023. 

\end{document}